\documentclass[9pt,twocolumn,twoside]{osajnl}

\usepackage{mathptmx,bm}

\journal{ol} 

\setboolean{shortarticle}{true}

\newcommand{\Tr}{\mathop{\mathrm{Tr}} \nolimits}
\newcommand{\op}[1]{\hat{#1}}

\newcommand{\ket}[1]{\left\vert #1 \right\rangle}

\newcommand{\openone}{\leavevmode\hbox{\small1\normalsize\kern-.33em1}}

\title{From polarization multipoles to higher-order coherences}

\author[1,2]{Aaron~Z.~Goldberg}
\author[3]{Andrei~B.~Klimov}
\author[4]{Hubert~deGuise}
\author[5,6]{Gerd~Leuchs}
\author[7,8,9]{Girish~S.~Agarwal}
\author[5,10,*]{Luis L. S\'anchez-Soto}

\affil[1]{National Research Council of Canada, Ottawa, Ontario K1A 0R6, Canada}
\affil[2]{Department of Physics, University of Toronto, Toronto, Ontario M5S 1A7, Canada}
\affil[3]{Departamento de F\'{\i}sica, Universidad de Guadalajara, 44420~Guadalajara, Jalisco, Mexico}
\affil[4]{Department of Physics, Lakehead University, Thunder Bay, Ontario P7B 5E1, Canada}
\affil[5]{Max-Planck-Institut f\"ur die Physik des Lichts, 91058 Erlangen,  Germany}
\affil[6]{Institute of Applied Physics, Russian Academy of Sciences, 603950 Nizhny Novgorod, Russia}
\affil[7]{Institute for Quantum Science and Engineering, Texas A\&M University, College Station, Texas 77843, USA}
\affil[8]{Department of Physics and Astronomy, Texas A\&M University, College Station, Texas 77843, USA}
\affil[9]{Department of Biological and Agricultural Engineering, Texas A\&M University, College Station, Texas 77843, USA}
\affil[10]{Departamento de \'Optica, Facultad de F\'{\i}sica, Universidad Complutense, 28040~Madrid,  Spain}
\affil[*]{Corresponding author: lsanchez@fis.ucm.es}

 \doi{\url{http://dx.doi.org/10.1364/XX.XX.XXXXXX}}

\begin{abstract}
We demonstrate that the multipoles associated with the density matrix are truly observable quantities that can be unambiguously determined from intensity moments. Given their correct transformation properties, these multipoles are the natural variables to deal with a number of problems in the quantum domain. In the case of polarization, the moments are measured after the light has passed through two quarter-wave plates, one half-wave plate, and a polarizing beam splitter for specific values of the angles of the waveplates. For more general two-mode problems, equivalent measurements can be performed.
\end{abstract}

\setboolean{displaycopyright}{true}

\begin{document}

\maketitle

Quantum information science capitalizes on the peculiar features of quantum mechanics to store, manipulate, and transfer information in ways that are {inaccessible} to any classical strategy.  Various  platforms have been proposed for implementing these tasks in practice. Among them, photons are one of the most effective probes to process information: they epitomize a naturally mobile and low-noise system with quantum-limited detection available~\cite{Wang:2020tr}.   

Photon-based quantum information employs suitable degrees of freedom, related to propagation direction (path encoding), light spatial structure (orbital angular momentum encoding), and time (time-bin encoding)~\cite{Flamini:2018vh,Slussarenko:2019vx}. Yet polarization is the degree of freedom most frequently used to encode the relevant information.  Its ubiquitous presence has been further enhanced by recent advances in entanglement generation, manipulation, and distribution and by the development of experimental frameworks for exploiting polarization using integrated devices~\cite{Wang:2020aa}.

In the continuous-variable regime, polarization is mostly utilized for the generation of nonclassical light, a basic resource for quantum information processing applications. Polarization squeezing~\cite{Chirkin:1993dz,Korolkova:2002fu}, which has been observed in numerous experiments~\cite{Bowen:2002kx,Heersink:2003aa,Dong:2007fu,Shalm:2009mi}, is perhaps the most tantalizing illustration. Stokes polarimetry~\cite{Goldberg:2019vj} is the measurement method of choice in this area; in fact, complete tomographic schemes are available in this case~\cite{Karassiov:2004xw} and have been implemented in the laboratory~\cite{Marquardt:2007bh,Muller:2012ys}.

In the discrete-variable regime of single, or few, photons, one treats the system in terms of multiphoton two-mode states. As a result, the polarization can be determined from correlation functions~\cite{White:1999fk,James:2001vn,Barbieri:2003ij,Bogdanov:2004bs,Adamson:2010ys,Altepeter:2011ly}. Since the Hilbert spaces involved have small dimensions, the state reconstruction can be efficiently achieved.

Irrespective of the regime, there is nowadays a general consensus that grasping all the intriguing effects emerging in the quantum world requires a proper understanding of higher-order polarization fluctuations~\cite{Goldberg:2021tx}. This is in the same vein as what happens in coherence theory, where a hierarchy of correlation functions is required to fully characterize the field state~\cite{Mandel:1995qy}. 

However, we should always bear in mind that polarization has a natural SU(2) invariance~\cite{Karassiov:1993lq}, by which we mean that two field states connected by an SU(2) transformation are equivalent concerning polarization statistics. Put differently, physics should be independent of the polarization basis chosen. Situations where such an invariance is not explicit can be a source of confusion.

An elegant way to incorporate this SU(2) invariance is by resorting to the notion of state multipoles~\cite{Fano:1959ly,Blum:1981rb}. Their standard formulation is in terms of the angular momentum formalism and they correspond to the successive moments of the generators (in our case, the Stokes variables): the dipole term is the first-order moment and can thus be identified with the classical picture, whereas the other multipoles account for higher-order fluctuations we want to examine. 

The two-mode formulation clearly reveals the vector nature of the field and the quantum excitations should appear in terms of multipoles. Since the modes are connected to Stokes variables via the time-honoured Jordan-Schwinger map~\cite{Jordan:1935aa,Schwinger:1965kx}, one is naturally led to examine the aspect of the state multipoles in the two-mode picture. This is precisely the problem we address in this Letter.  {The approach} is by no means restricted to polarization, but encompasses many other instances, such as, e.g., strongly correlated systems~\cite{Auerbach:1988ta}, Bose-Einstein condensates~\cite{Ma:2011xd}, and Gaussian-Schell beams~\cite{Sundar:1995tu} (see Ref.~\cite{Chaturvedi:2006vn} for a complete review). {We emphasize that the mode amplitudes are precisely the experimentally accessible variables in most of those examples.} 

{We work} out a new compact expression for the state multipoles that allows us to reinterpret them in terms of field correlation functions. This allows one to determine these multipoles from simple measurements.  More importantly, since the pioneering contributions of Wolf~\cite{Wolf:2007aa},  we know that there is an intimate relationship between polarization properties of a random beam and its coherence properties at the classical level. Our results extend such a connection to arbitrary higher orders in the quantum domain. 

Let us set the stage for our discussion. We will be considering a monochromatic field specified by two operators $\op{a}_{H}$ and $\op{a}_{V}$, representing the complex amplitudes in two linearly polarized orthogonal modes that we indicate as horizontal ($H$) and vertical ($V$), respectively. These operators obey the bosonic commutation rules {$[\op{a}_{\sigma}, \op{a}_{\sigma^{\prime}}^{\dagger} ] = \delta_{\sigma \sigma^{\prime}}$  ($\sigma, \sigma^\prime \in \{ H, V \})$,} with the superscript  $\dagger$ standing for the Hermitian adjoint  and $\hbar = 1$ throughout.  The operator $ \op{N} = \op{a}^{\dagger}_{H} \op{a}_{H} +   \op{a}^{\dagger}_{V} \op{a}_{V}$ represents then the total number of photons.  

In classical optics, the polarization state is portrayed as a point on the Poincar\'e sphere, with  coordinates given by the Stokes parameters. As the total intensity is a well-defined quantity, the Poincar\'e sphere has a distinct radius (equal to the intensity).  In quantum optics, however, fluctuations in the number of photons are unavoidable and we are forced to consider instead a three-dimensional Poincar\'e space that can be envisioned as a set of nested spheres with radii proportional to the different total photon numbers that contribute to the state.

In this onion-like picture, each shell (also known as a Fock layer~\cite{Muller:2016aa}) has to be addressed independently.  This can be stressed if the Fock states $\{ |n_H{\rangle\otimes|} n_V \rangle \} $, which comprise the standard basis for two-mode fields, are relabeled as {$\ket{S, m } \equiv \ket{n_H = S + m}\otimes\ket{ n_V = S - m}$}. Equivalently, one has $S = \tfrac{1}{2} (n_{H} + n_{V})$ and $m =\tfrac{1}{2} (n_{H} - n_{V})$ in a standard angular-momentum notation. In this way, for each fixed $S$ (i.e., fixed number of photons), $m$ runs from $-S$ to $S$ and these states span a $(2S+1)$-dimensional subspace $\mathcal{H}_{S}$ that corresponds to a single Fock layer of \textit{spin} $S$.  

The density matrix $\op{\varrho}$ of a general two-mode state spans many Fock layers. However, expanding $\op{\varrho}$ in the basis {$\{ \ket{S,m} \}$} is not a convenient choice because it does not explicitly reflect the aforementioned SU(2) invariance. It is more convenient to use instead the so-called irreducible tensor operators $\hat{T}_{Kq}$, which are tailored to transform covariantly under SU(2); viz.~\cite{Fano:1959ly,Blum:1981rb},
\begin{equation}
  \op{R} ( \phi, \theta, \psi) \; \op{T}_{Kq}  \;
  \op{R}^{\dagger} ( \phi, \theta, \psi) =
  \sum_{q^{\prime}}  D_{q^{\prime} q}^{K} ( \phi, \theta, \psi) \;
  \op{T}_{Kq^{\prime}}  \, ,
  \label{eq:sym3}
\end{equation}
where $\op{R}$ is the operator representing a rotation of Euler angles $ (\phi, \theta, \psi)$ and $D_{q^{\prime} q}^{K} ( \phi, \theta, \psi) $ stands for the Wigner $D$-matrix~\cite{Varshalovich:1988xy}, which encodes the matrix elements of $\op{R} $ in the basis $ \{ |S,  m \rangle \}$; i.e., $D_{m^{\prime} m}^{S} ( \phi, \theta, \psi)  = \langle S,m^{\prime} |\op{R} ( \phi, \theta, \psi) |S,m \rangle$.   

The standard construction of these tensors is in terms of the angular momentum basis. However, its two-mode counterpart is, surprisingly, missing. To {overcome this lack}, we {define a set of tensor operators different from the usual construction:}
\begin{equation}
\label{eq:tensors}
\op{T}_{Kq} = \frac{\hat{a}_{H}^{\dagger K+q} \hat{a}_{V}^{\dagger K-q}}
{\sqrt{(K+q)!(K-q)!}} \, ,
\end{equation}
with $K$ taking any integer or half-integer value and $- K \leq  q \leq K${. These new operators} indeed transform as required by \eqref{eq:sym3} {and} will play a crucial role in what follows. 

Notice that we have chosen the numerical factors in \eqref{eq:tensors} so as to fulfill {$\op{T}_{Kq} |\mathrm{vac} \rangle = | K,q \rangle$, where $|\mathrm{vac} \rangle {=\ket{0}\otimes\ket{0}}$ is the two-mode vacuum. This is tantamount} to ensuring that the tensors connect two Fock layers whose total numbers of photons differ by $2K$. {The definition of the Wigner $D$-matrices through $|K,q^\prime\rangle=\sum_q D_{q^\prime q}^K |K,q\rangle$ and that $\op{R}^\dagger ( \phi, \theta, \psi)   |\mathrm{vac}\rangle=|\mathrm{vac}\rangle$ immediately validate the covariant nature of our chosen operators $\op{T}_{Kq}$.} With this choice, for a fixed $S$, we have that {$\Tr ( \hat{T}_{Kq}  \hat{T}_{K^\prime q^\prime}^{\dagger}  ) = C(2K+2S+1, 2K+1)  \delta_{KK^{\prime}} \delta_{qq^{\prime}}$, where $C({n},{k})$ is a binomial coefficient.} This has to be taken into consideration when expanding observables in terms of these tensors. 

\begin{figure}[t]
\centering{\includegraphics[width=0.85\columnwidth]{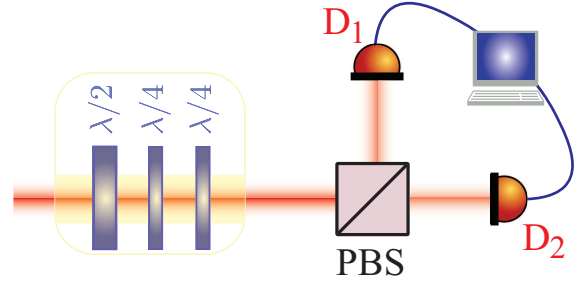}}
\caption{Sketch of the setup with an SU(2) gadget, composed of two quarter-wave plates and  one half-wave plate whose orientation angles may be adjusted, and a polarizing beam splitter (PBS) that separates components $H$ and $V$. We assume that, at the detectors $D_{1}$ and $D_{2}$, the $K$th-order intensity moment is measured.}
\label{fig:1}
\end{figure}

Next, let us consider the setup sketched in Fig.~\ref{fig:1}. The light beam first encounters a half-wave plate and two quarter-wave plates, all mounted coaxially. This wave plate arrangement constitutes a universal SU(2) gadget for polarized light: {to realize a given $\op{R} ( \phi, \theta, \psi)$ one simply has to rotate these plates about the common axis to angular positions characteristic of the element}~\cite{Simon:1989uf}. Recall that SU(2) transformations do not change the number of photons. For each fixed value of the Euler angles $( \phi, \theta, \psi)$, the setup measures the intensity moments {of the rotated state}, as suggested in Ref.~\cite{Schilling:2010aa} and implemented experimentally in Ref.~\cite{Israel:2012vd}. 

To analyze such measurements in a completely SU(2)-invariant form, we {restrict our attention to a single Fock layer. When the photon number is indefinite, we parse the state in Fock layers and apply the same technique  to each layer~\cite{Muller:2016aa}. We first note that} the matrix
\begin{equation}
    G_{qq^{\prime}}^{K} = \Tr ( 
    \op{\varrho} \op{T}_{Kq} \, \op{T}_{Kq^{\prime}}^{\dagger}) 
\end{equation}  
does contain (apart from an unessential global constant) all Glauber-ordered $K$th field correlations~\cite{Mandel:1991aa}, which are indispensable for a complete state characterization. For example, the state $|\psi\rangle=\sum_{q} \psi_q| K,q \rangle$ has correlations $G^K_{q,q^\prime}=\psi_q^*\psi_{q^\prime}$. In consequence, from this perspective, the intensity moments measured by the setup, which enacts $\op{\varrho} \mapsto \op{R}( \phi, \theta, \psi) \; \op{\varrho} \, \op{R}^\dagger( \phi, \theta, \psi)$ before splitting and measuring the two polarization modes, can be appropriately described by
\begin{align}
{I}_{Kq}( \theta, \phi ) & =  \Tr [ \op{\varrho} \, \hat{R} ( \phi, \theta, \psi)  \, 
\op{T}_{Kq}  \op{T}_{Kq}^{\dagger} \, \hat{R}^{\dagger} ( \phi, \theta, \psi)  ] \nonumber \\
& = \sum_{q^{\prime} q^{\prime \prime}} 
G_{q^{\prime \prime} q^{\prime}}^{K} \;
D_{q^{\prime \prime}  q}^{K} ( \phi, \theta, \psi)  D_{q^{\prime} q}^{K \ast}( \phi, \theta, \psi)  \, . 
\end{align}
If we take into account that  $D_{q^{\prime} q}^{K \ast} = (-1)^{q-q^{\prime}} D_{- q^{\prime} \, -q}^{K}$ and the expression for the product of two $D$-functions~\cite{Varshalovich:1988xy},  it turns out that 
\begin{align}
\label{eq:Ikqfin}
{I}_{Kq}(\theta, \phi ) & =   \sum_{{L}=0}^{2K} \sqrt{\frac{4\pi }{2{L} + 1}}  \sum_{q^{\prime}, q^{\prime \prime}}  
(-1)^{q-q^{\prime}}  C_{Kq, K -q}^{{L}0}  G_{q^{\prime \prime} q^{\prime}}^{K}  \nonumber \\
& \times  \sum_{m} C_{Kq^{\prime \prime}, K -q^{\prime}}^{{L}m} \, Y_{{L}m}^{\ast }(\theta , \phi ) \, ,
\end{align}
where $C_{{L}_{1}m_{1}, {L}_{2} m_{2}}^{{L}m}$ is a standard Clebsch-Gordan coefficient and we have taken into account that $D_{m0}^{{L}} (\phi, \theta, \psi ) = [{4\pi}/{(2{L}+1)}]^{1/2} Y_{{L}m}^{\ast} (\theta, \phi)$, where $Y_{{L}m}^{\ast}$ denotes the corresponding spherical harmonic~\cite{Varshalovich:1988xy}. To lighten notation, we omit everywhere the limits on summations over the third component of an angular momentum, as they always extend to their natural domain; e.g., $- {L} \le m \le {L}$. {This is the sense in which we have multipole moments: we are looking to find the expansion coefficients of the intensity moments in terms of spherical harmonics, which are uniquely determined by the correlations $G_{qq^\prime}^K$.}

This remarkable expression clearly shows that the intensity moments depend exclusively on the correlation functions $G_{q^{\prime \prime} q^{\prime}}^{K}$ and thus on the corresponding multipoles. In order to be truly useful, we should be able to invert this formula so as to express $G_{q^{\prime \prime} q^{\prime}}^{K}$ in terms of the measured $I_{Kq}$, such that all of the field correlations can be determined from simple intensity measurements~\cite{Mukunda:1966wf}. This can be accomplished by multiplying both sides of \eqref{eq:Ikqfin} by $Y_{{L}^{\prime }m^{\prime }}(\theta, \phi)$ and integrating over all angles
\begin{eqnarray}
& \displaystyle \int d\Omega \;  {I}_{Kq}( \theta, \phi ) Y_{{L}^{\prime}m^{\prime}}(\theta , \phi ) = \sqrt{\frac{4\pi }{2{L}^{\prime }+1}} C_{Kq,K-q}^{{L}^{\prime}0} & \nonumber \\
& \displaystyle  \times \sum_{q^{\prime},q^{\prime \prime }} 
(-1)^{q-q^{\prime }} \, C_{Kq^{\prime \prime },K-q^{\prime }}^{{L}^{\prime }m^{\prime
}} \, G_{q^{\prime \prime}q^{\prime}}^{K} \, ,& 
\end{eqnarray} 
where $d \Omega = \sin \theta \, d \theta d \phi$ is the invariant measure on the sphere. Finally, we multiply by $C_{Kp^{\prime \prime},K-p^{\prime}}^{{L}^{\prime}m^{\prime}}$, sum over ${L}^\prime$ and $m^\prime$, and use the orthogonality relations of the Clebsch-Gordan coefficients; the inversion we are looking for reads
\begin{align}
\label{eq:Ginv}
G_{q^{\prime \prime} q^{\prime }}^{K} & = (-1)^{q-q^{\prime }} \sum_{{L}=0}^{2K}\sum_{m} \sqrt{\frac{2{L}+1}{4\pi }}%
\left( C_{Kq,K-q}^{{L}0}\right) ^{-1} \nonumber \\
& \times \int d\Omega \; {I}_{Kq}(\theta, \phi) \, Y_{{L}m}(\theta, \phi ) 
C_{Kq^{\prime \prime },K-q^{\prime }}^{{L}m} \, ,
\end{align}
where any value of the index $q$ can be taken. To determine \emph{all} of the correlations and \emph{all} of the intensity moments, only a \emph{single} type of projection needs to be measured: by simply varying the rotations imparted on the state by the waveplates in Fig.~\ref{fig:1} and repeating the measurement, all of the polarization information can be recovered. When the initial state has an exact number of photons (i.e., lies in a single Fock layer), either one of the two detectors in the setup will suffice for determining ${I}_{Kq}( \theta, \phi)$, as can be directly verified by choosing $q=\pm K$.

The inversion in \eqref{eq:Ginv}, although simple and elegant, requires a continuum sampling of the angles $\Omega$, which is a serious disadvantage from an experimental point of view. On physical grounds, one would expect that the $K$th-order correlations $G_{q^{\prime \prime} q^{\prime}}^{K}$ can be decomposed in terms of multipoles of order ${L}=0,\cdots,2K$, which can each be determined by performing $2{L}+1$ independent measurements {to capture the angular information in the $2{L}+1$ linearly independent spherical harmonics $Y_{{L}m}(\theta, \phi)$ of order ${L}$}; i.e., measure $I_{Kq}(\theta_{i}, \phi_{i})$ for $\sum_{{L}=0}^{2K}(2{L}+1)=(2K+1)^2$ values of $i$ in total. 

To formalize this idea, we first note that, using the property $\sum_{q}  C_{Kq, K -q}^{S0} \; C_{Kq, K -q}^{L0} = \delta_{SL}$, \eqref{eq:Ikqfin} can be recast as 
\begin{align}
\label{eq:resu}
\sum_{q} C_{Kq, K -q}^{{L}0} \, {I}_{Kq}(\theta, \phi )  & =  
\sqrt{\frac{4\pi }{2{L} +1}} \sum_{q^{\prime}, q^{\prime \prime}}  
(-1)^{q-q^{\prime \prime}} G_{q^{\prime \prime} q^{\prime}}^{K}  
\nonumber \\
& \times  \sum_{m=-{L}}^{{L}} C_{Kq^{\prime}, K -q^{\prime \prime}}^{{L}m} Y_{{L}m}^{\ast }(\theta, \phi ) \, .
\end{align}
If we define 
\begin{align}
\widetilde{I}_{{L}} (\theta,  \phi ) & =  \sum_{q} C_{Kq, K -q}^{{L}0} \;
{I}_{Kq}(\theta , \phi )\, ,\ \nonumber \\
& \\
\widetilde{G}_{{L}}^{(m)} & =  \sum_{q^{\prime}, q^{\prime \prime}}  
(-1)^{q-q^{\prime}} C_{Kq^{\prime \prime}, K -q^{\prime}}^{{L}m}\,  G_{q^{\prime \prime}q^{\prime}}^{K} \, ,\nonumber  
\end{align}
the result (\ref{eq:resu}) can be finally written in the compact form 
\begin{equation}
\label{eq:guay}
    \widetilde{I}_{{L}} (\theta, \phi )=\sqrt{\frac{4\pi}{2{L}+1}}\sum_{m} Y_{{L}m}^{\ast }(\theta, \phi ) \; \widetilde{G}_{{L}}^{(m)} \, .
\end{equation}
Here, $\widetilde{I}_{{L}} (\theta, \phi  )$ and $\widetilde{G}_{{L}}^{(m)}$ are Schur transforms of the intensity moments ${I}_{Kq}(\theta, \phi  )$ and $G_{q^{\prime \prime}q^\prime}^K$, respectively, which can readily be inverted by again using the orthonormality relations among the Clebsch-Gordan coefficients. We have suppressed the dependence of the transforms on $K$ because the inversion formula will be independent from $K$. This transform, sometimes referred to as a Clebsch-Gordan transform~\cite{Bacon:2006wg,Giacometti:2009tk}, maps the computational basis (in our case, the two-mode Fock basis {$\ket{n_H}\otimes\ket{n_V}$}) to the Schur basis (in our case, the $|{L},m \rangle$ basis). We now see that the multipole moments for expanding the intensity distribution in terms of spherical harmonics come from a Clebsch-Gordan transform of the correlation functions in the polarization state.

{The method then proceeds independently for each order: first, we measure the first-order intensity moments {$I_{1q}\left(\theta, \phi\right)$} in the directions of the three coordinate axes (or any other equivalent directions) so the resulting system in \eqref{eq:guay} can be immediately solved independently from $K$, getting
\begin{equation}
    \begin{pmatrix}
    \widetilde{G}_{1}^{(1)} \\
    \widetilde{G}_{1}^{(0)} \\
    \widetilde{G}_{1}^{(-1)}
    \end{pmatrix} =  \frac{1}{\sqrt{3}}  
    \begin{pmatrix}
    -1 & i & 0  \\
    0 & 0 & \sqrt{2} \\
    1 & i & 0
    \end{pmatrix}
    \begin{pmatrix}
    \widetilde{I}_{1} {(\mathbf{x})} \\
    \widetilde{I}_{1} {(\mathbf{y})} \\
    \widetilde{I}_{1} {(\mathbf{z})}
    \end{pmatrix} \, .
\end{equation}
This allows us to infer all of the first-order properties.}

Next, we proceed much in the same way: measure the second-order intensity moments in five optimal independent directions and solve the ensuing system. For example, we can choose the directions that maximize the minimum angle between the lines and thus in some sense spread the measurements over the Poincar\'e sphere as widely as possible~\cite{Hoz:2013om}.

For the $L$th moment, the inversion of the system can be achieved in a closed manner~\cite{Hoz:2013om}, again without the inversion depending on $K$ or the previous orders:
\begin{equation}
\widetilde{\mathbf{G}}_{{L}} = \frac{4\pi}{2{{L}}+1} \mathbf{P}_{{{L}}}^{-1} \; \mathbf{Y}_{{{L}}}^{\dagger}\; \widetilde{\mathbf{I}}_{{L}} \, .
\end{equation}
We have introduced the vectors $\widetilde{\mathbf{G}}_{{L}} = (\widetilde{G}_{{{L}}}^{(-L)}, \ldots, \widetilde{G}_{{{L}}}^{(L)})^{\top}$ and  $\widetilde{\mathbf{I}}_{{L}} = (\widetilde{I}_{{{L}}}(\theta_1,\phi_1), \ldots, \widetilde{I}_{{{L}}}(\theta_{2L+1},\phi_{2L+1}))^{\top}$ (the superscript $\top$ denoting the transpose) and, in addition, we have
\begin{equation}
[\mathbf{Y}_{{{L}}}]_{jk}  = Y_{{{L}}k} (\theta_j, \phi_j) \, , 
\qquad \qquad
[\mathbf{P}_{{{L}}}]_{jk}  = P_{{{L}}}(\chi_{jk})  \, ,
\end{equation}
 with $\cos \chi_{jk} =\cos \theta_{j}\cos \theta _{k}+\sin \theta _{j}\sin \theta _{k}\sin (\phi _{j}-\phi_{k})$, and $P_{{{L}}}(x)$ representing Legendre polynomials. That is our central result: we have provided an explicit method for determining all of the intensity correlations $G_{qq^\prime}^K$ and thereby the multipole moments of the intensity distribution from a set of $(2K+1)^2$ identical measurements on the polarization state rotated to $(2K+1)^2$ different orientations. These can then be readily determined for all orders $K$, which cannot exceed $S$ for states with fixed total number of photons $2S$.

 Choosing the appropriate directions is, in general, a tricky question if one wants to be sure about the linear independence, but it has been thoroughly studied~\cite{Newton:1968ve,Filippov:2010kx}. In practice, methods such as maximum likelihood are much more efficient at handling that inversion~\cite{lnp:2004uq}.

Nature conspires to make a gadget that measures intensity moments more useful than one that  directly measures some correlation $G_{qq^\prime}^K$ for $q\neq q^\prime$. In contrast to the intensity correlations given by ${I}_{Kq}(\phi, \theta, \psi)$, a gadget that measures the correlations $G_{qq^\prime}^K$ for some particular $q\neq q^\prime$ would be insufficient for determining all of the Glauber-ordered $K$th field correlations, as the correlation is not a sum of $(2K+1)^2$ linearly independent functions of $(\phi, \theta, \psi)$.

An extension of these results could be obtained if one had access to a gadget that measures a correlation between different Fock layers, such as some particular $G_{qq}^{KK^\prime} = \Tr ( \op{\varrho} \op{T}_{Kq} \op{T}^{\dagger}_{K^\prime q} )$. Then, the measurements $\langle \op{R} (\phi, \theta, \psi)  \,  \op{T}_{Kq} \op{T}^{\dagger}_{K^\prime q} \, \hat{R}^{\dagger} (\phi, \theta, \psi )  \rangle$ over an appropriate set of angles $(\phi, \theta, \psi)$ could be inverted to produce all of the correlations $G_{qq^\prime}^{KK^\prime}$ for that particular pair of $K$ and $K^\prime$. This sort of analysis is only possible because we expressed our tensors as two-mode operators.

As the simplest example, a gadget that replaces the detectors in Fig.~\ref{fig:1} with ones that measures the correlation $\langle \op{a}_H^\dagger\op{a}_V^\dagger \rangle$ could be used to determine all of the correlations $\langle \op{a}_H^\dagger\op{a}_V^\dagger \rangle$, $\langle \op{a}_H^{\dagger 2} \rangle$, and $\langle \op{a}_V^{\dagger 2} \rangle$, which could be achieved using photon addition. Again, this would be more useful than a gadget that measures a particular correlation $G_{qq^\prime}^{KK^\prime}$ for $q\neq q^\prime$, as the latter does not contain sufficient information to reproduce all of the desired correlations.

In summary, we have worked out a detailed formulation of the machinery of SU(2) irreducible tensors for two-mode fields. Since, in many instances, the experimentally accessible observables are precisely the mode amplitudes, our formalism should play a major role in a proper analysis of these kinds of experiments. As a confirmation, we have demonstrated that the state multipoles are nothing but the Glauber correlation functions when expressed in the mode picture. This confirms a conspicuous relation between polarization and coherence in the fully quantum realm.

\begin{backmatter}
\bmsection{Funding}
Natural Sciences and Engineering Research Council of Canada (NSERC), Consejo Nacional de Ciencia y Tecnologia (254127), European Union's Horizon 2020 Research and Innovation program (ApresSF and STORMYTUNE), Air Force Office of Scientific Research (FA9550-20-1-0366), WELCH Foundation (A-1943), Ministerio de Ciencia e Innovaci{\'o}n (PGC2018-099183-B-I00).

\bmsection{Disclosures}
The authors declare no conflicts of interest

\bmsection{Data Availability Statement}
All data generated or analyzed during this study are included in this published article. 
\end{backmatter}


\begin{thebibliography}{10}
\newcommand{\enquote}[1]{``#1''}

\bibitem{Wang:2020tr}
J.~Wang, F.~Sciarrino, A.~Laing, and M.~G. Thompson, \enquote{Integrated
  photonic quantum technologies,} {\protect\JournalTitle{Nat. Photonics}}
  \textbf{14}, 273--284 (2020).

\bibitem{Flamini:2018vh}
F.~Flamini, N.~Spagnolo, and F.~Sciarrino, \enquote{Photonic quantum
  information processing: a review,} {\protect\JournalTitle{Rep. Prog. Phys.}}
  \textbf{82}, 016001 (2018).

\bibitem{Slussarenko:2019vx}
S.~Slussarenko and G.~J. Pryde, \enquote{Photonic quantum information
  processing: A concise review,} {\protect\JournalTitle{Appl. Phys. Rev.}}
  \textbf{6}, 041303 (2019).

\bibitem{Wang:2020aa}
J.~Wang, F.~Sciarrino, A.~Laing, and M.~G. Thompson, \enquote{Integrated
  photonic quantum technologies,} {\protect\JournalTitle{Nat. Photonics}}
  \textbf{14}, 273--284 (2020).

\bibitem{Chirkin:1993dz}
A.~S. Chirkin, A.~A. Orlov, and D.~Y. Parashchuk, \enquote{Quantum theory of
  two-mode interactions in optically anisotropic media with cubic
  nonlinearities: Generation of quadrature- and polarization-squeezed light,}
  {\protect\JournalTitle{Quantum Electron.}} \textbf{23}, 870--874 (1993).

\bibitem{Korolkova:2002fu}
N.~Korolkova, G.~Leuchs, R.~Loudon, T.~C. Ralph, and C.~Silberhorn,
  \enquote{Polarization squeezing and continuous-variable polarization
  entanglement,} {\protect\JournalTitle{Phys. Rev. A}} \textbf{65}, 052306
  (2002).

\bibitem{Bowen:2002kx}
W.~P. Bowen, R.~Schnabel, H.-A. Bachor, and P.~K. Lam, \enquote{Polarization
  squeezing of continuous variable {S}tokes parameters,}
  {\protect\JournalTitle{Phys. Rev. Lett.}} \textbf{88}, 093601 (2002).

\bibitem{Heersink:2003aa}
J.~Heersink, T.~Gaber, S.~Lorenz, O.~Gl{\"o}ckl, N.~Korolkova, and G.~Leuchs,
  \enquote{Polarization squeezing of intense pulses with a fiber-optic sagnac
  interferometer,} {\protect\JournalTitle{Physical Review A}} \textbf{68},
  013815-- (2003).

\bibitem{Dong:2007fu}
R.~Dong, J.~Heersink, J.-I. Yoshikawa, O.~Gl{\"o}ckl, U.~L. Andersen, and
  G.~Leuchs, \enquote{An efficient source of continuous variable polarization
  entanglement,} {\protect\JournalTitle{New J. Phys.}} \textbf{9}, 410 (2007).

\bibitem{Shalm:2009mi}
L.~K. Shalm, A.~R.~B. A., and A.~M. Steinberg, \enquote{Squeezing and
  over-squeezing of triphotons,} {\protect\JournalTitle{Nature}} \textbf{457},
  67--70 (2009).

\bibitem{Goldberg:2019vj}
A.~Z. {Goldberg}, \enquote{{Quantum theory of polarimetry: From quantum
  operations to Mueller matrices},} {\protect\JournalTitle{Phys. Rev.
  Research}} \textbf{2}, 023038 (2019).

\bibitem{Karassiov:2004xw}
V.~P. Karassiov and A.~V. Masalov, \enquote{The method of polarization
  tomography of radiation in quantum optics,} {\protect\JournalTitle{JETP}}
  \textbf{99}, 51--60 (2004).

\bibitem{Marquardt:2007bh}
C.~Marquardt, J.~Heersink, R.~Dong, M.~V. Chekhova, A.~B. Klimov, L.~L.
  S{\'a}nchez-Soto, U.~L. Andersen, and G.~Leuchs, \enquote{Quantum
  reconstruction of an intense polarization squeezed optical state,}
  {\protect\JournalTitle{Phys. Rev. Lett.}} \textbf{99}, 220401 (2007).

\bibitem{Muller:2012ys}
C.~R. M{\"u}ller, B.~Stoklasa, C.~Peuntinger, C.~Gabriel, J.~{\v R}eh{\'a}{\v
  c}ek, Z.~Hradil, A.~B. Klimov, G.~Leuchs, C.~Marquardt, and L.~L.
  S{\'a}nchez-Soto, \enquote{Quantum polarization tomography of bright squeezed
  light,} {\protect\JournalTitle{New J. Phys.}} \textbf{14}, 085002 (2012).

\bibitem{White:1999fk}
A.~G. White, D.~F.~V. James, P.~H. Eberhard, and P.~G. Kwiat,
  \enquote{Nonmaximally entangled states: Production, characterization, and
  utilization,} {\protect\JournalTitle{Phys. Rev. Lett.}} \textbf{83},
  3103--3107 (1999).

\bibitem{James:2001vn}
D.~F.~V. James, P.~G. Kwiat, W.~J. Munro, and A.~G. White, \enquote{Measurement
  of qubits,} {\protect\JournalTitle{Phys. Rev. A}} \textbf{64}, 052312 (2001).

\bibitem{Barbieri:2003ij}
M.~Barbieri, F.~De~Martini, G.~Di~Nepi, P.~Mataloni, G.~M. D'Ariano, and
  C.~Macchiavello, \enquote{Detection of entanglement with polarized photons:
  Experimental realization of an entanglement witness,}
  {\protect\JournalTitle{Phys. Rev. Lett.}} \textbf{91}, 227901 (2003).

\bibitem{Bogdanov:2004bs}
Y.~I. Bogdanov, M.~V. Chekhova, S.~P. Kulik, G.~A. Maslennikov, A.~A. Zhukov.,
  C.~H. Oh, and M.~K. Tey, \enquote{Qutrit state engineering with biphotons,}
  {\protect\JournalTitle{Phys. Rev. Lett.}} \textbf{93}, 230503 (2004).

\bibitem{Adamson:2010ys}
R.~B.~A. Adamson and A.~M. Steinberg, \enquote{Improving quantum state
  estimation with mutually unbiased bases,} {\protect\JournalTitle{Phys. Rev.
  Lett.}} \textbf{105}, 030406 (2010).

\bibitem{Altepeter:2011ly}
J.~B. Altepeter, N.~N. Oza, M.~Medi, E.~R. Jeffrey, and P.~Kumar,
  \enquote{Entangled photon polarimetry,} {\protect\JournalTitle{Opt. Express}}
  \textbf{19}, 26011--26016 (2011).

\bibitem{Goldberg:2021tx}
A.~Z. Goldberg, P.~de~la Hoz, G.~Bj{\"o}rk, A.~B. Klimov, M.~Grassl, G.~Leuchs,
  and L.~L. S{\'a}nchez-Soto, \enquote{Quantum concepts in optical
  polarization,} {\protect\JournalTitle{Adv. Opt. Photon.}} \textbf{13}, 1--73
  (2021).

\bibitem{Mandel:1995qy}
L.~Mandel and E.~Wolf, \emph{Optical Coherence and Quantum Optics} (Cambridge
  University, 1995).

\bibitem{Karassiov:1993lq}
V.~P. Karassiov, \enquote{Polarization structure of quantum light fields: a new
  insight. {I}. general outlook,} {\protect\JournalTitle{J. Phys. A}}
  \textbf{26}, 4345--4354 (1993).

\bibitem{Fano:1959ly}
U.~Fano and G.~Racah, \emph{Irreducible Tensorial Sets} (Academic, 1959).

\bibitem{Blum:1981rb}
K.~Blum, \emph{Density Matrix Theory and Applications} (Plenum, 1981).

\bibitem{Jordan:1935aa}
P.~Jordan, \enquote{Der {Z}usammenhang der symmetrischen und linearen {G}ruppen
  und das {M}ehrk{\"o}rperproblem,} {\protect\JournalTitle{Z. Phys.}}
  \textbf{94}, 531--535 (1935).

\bibitem{Schwinger:1965kx}
J.~Schwinger, \enquote{On angular momentum,} in \emph{Quantum Theory of Angular
  Momentum,}  L.~C. Biedenharn and H.~Dam, eds. (Academic, New York, 1965).

\bibitem{Auerbach:1988ta}
A.~Auerbach and D.~P. Arovas, \enquote{Spin dynamics in the square-lattice
  antiferromagnet,} {\protect\JournalTitle{Phys. Rev. Lett.}} \textbf{61},
  617--620 (1988).

\bibitem{Ma:2011xd}
J.~Ma, X.~Wang, C.~P. Sun, and F.~Nori, \enquote{Quantum spin squeezing,}
  {\protect\JournalTitle{Phys. Rep.}} \textbf{509}, 89--165 (2011).

\bibitem{Sundar:1995tu}
K.~Sundar, N.~Mukunda, and R.~Simon, \enquote{Coherent-mode decomposition of
  general anisotropic gaussian schell-model beams,} {\protect\JournalTitle{J.
  Opt. Soc. Am. A}} \textbf{12}, 560--569 (1995).

\bibitem{Chaturvedi:2006vn}
S.~Chaturvedi, G.~Marmo, and N.~Mukunda, \enquote{The {S}chwinger
  representation of a group: concept and applications,}
  {\protect\JournalTitle{Rev. Math. Phys.}} \textbf{18}, 887--912 (2006).

\bibitem{Wolf:2007aa}
E.~Wolf, \emph{Introduction to the Theory of Coherence and Polarization of
  Light} (Cambridge University, 2007).

\bibitem{Muller:2016aa}
C.~R. M{\"u}ller, L.~S. Madsen, A.~B. Klimov, L.~L. S{\'a}nchez-Soto,
  G.~Leuchs, C.~Marquardt, and U.~L. Andersen, \enquote{Parsing polarization
  squeezing into {F}ock layers,} {\protect\JournalTitle{Phys. Rev. A}}
  \textbf{93}, 033816 (2016).

\bibitem{Varshalovich:1988xy}
D.~A. Varshalovich, A.~N. Moskalev, and V.~K. Khersonski, \emph{{Quantum Theory
  of Angular Momentum}} (World Scientific, 1988).

\bibitem{Simon:1989uf}
R.~Simon and N.~Mukunda, \enquote{Universal su(2) gadget for polarization
  optics,} {\protect\JournalTitle{Phys. Lett. A}} \textbf{138}, 474--480
  (1989).

\bibitem{Schilling:2010aa}
U.~Schilling, J.~von Zanthier, and G.~S. Agarwal, \enquote{Measuring
  arbitrary-order coherences: Tomography of single-mode multiphoton
  polarization-entangled states,} {\protect\JournalTitle{Phys. Rev. A}}
  \textbf{81}, 013826-- (2010).

\bibitem{Israel:2012vd}
Y.~Israel, I.~Afek, S.~Rosen, O.~Ambar, and Y.~Silberberg,
  \enquote{Experimental tomography of noon states with large photon numbers,}
  {\protect\JournalTitle{Phys. Rev. A}} \textbf{85}, 022115 (2012).

\bibitem{Mandel:1991aa}
L.~Mandel, \enquote{Coherence and indistinguishability,}
  {\protect\JournalTitle{Opt. Lett.}} \textbf{16}, 1882--1883 (1991).

\bibitem{Mukunda:1966wf}
N.~Mukunda and T.~F. Jordan, \enquote{Determination of optical field
  correlations from photon counts,} {\protect\JournalTitle{J. Math. Phys.}}
  \textbf{7}, 849--853 (1966).

\bibitem{Bacon:2006wg}
D.~Bacon, I.~L. Chuang, and A.~W. Harrow, \enquote{Efficient quantum circuits
  for schur and {Clebsch-Gordan} transforms,} {\protect\JournalTitle{Phys. Rev.
  Lett.}} \textbf{97}, 170502 (2006).

\bibitem{Giacometti:2009tk}
A.~Giacometti, J.~Lado, F.~Largo, G.~Pastore, and F.~Sciortino, \enquote{Phase
  diagram and structural properties of a simple model for one-patch particles,}
  {\protect\JournalTitle{J. Chem. Phys.}} \textbf{131}, 174114 (2009).

\bibitem{Hoz:2013om}
P.~de~la Hoz, A.~B. Klimov, G.~Bj{\"o}rk, Y.~H. Kim, C.~M{\"u}ller,
  C.~Marquardt, G.~Leuchs, and L.~L. S{\'a}nchez-Soto, \enquote{Multipolar
  hierarchy of efficient quantum polarization measures,}
  {\protect\JournalTitle{Phys. Rev. A}} \textbf{88}, 063803 (2013).

\bibitem{Newton:1968ve}
R.~G. Newton and B.-l. Young, \enquote{Measurability of the spin density
  matrix,} {\protect\JournalTitle{Ann. Phys.}} \textbf{49}, 393--402 (1968).

\bibitem{Filippov:2010kx}
S.~N. Filippov and V.~I. Man'ko, \enquote{Inverse spin-s portrait and
  representation of qudit states by single probability vectors,}
  {\protect\JournalTitle{J. Russ. Las. Res.}} \textbf{31}, 32--54 (2010).

\bibitem{lnp:2004uq}
M.~G.~A. Paris and J.~\v{R}eh\'a\v{c}ek, eds., \emph{Quantum State Estimation},
  vol. 649 of \emph{Lecture Notes in Physics} (Springer, 2004).

\end{thebibliography}

\newpage



\end{document}